%% file: CMEshockpaper.tex
\input{psfig.sty}

\documentclass{aa} 

\usepackage{times}
\usepackage{natbib}
\usepackage{tabularx}
\usepackage{graphicx}
\usepackage{wasysym}

\begin{document}

\title{Modeling magnetohydrodynamics and non equilibrium SoHO/UVCS line emission of CME shocks}
\author{P.Pagano\inst{1,2,3} \and J.C. Raymond\inst{2} \and F.Reale\inst{1,3} \and S.Orlando\inst{3}}
\institute{{Dipartimento di Scienze Fisiche ed Astronomiche, Sezione di Astronomia, Universit\`a di Palermo, Piazza del Parlamento 1, 90134 Palermo, Italy}
\and
{Harvard-Smithsonian Center for Astrophysics, 60 Garden Street, Cambridge, MA 02138}
\and
{INAF - Osservatorio Astronomico di Palermo, Piazza del Parlamento 1, 90134 Palermo, Italy}}
   
\newcommand{\btx}{\textsc{Bib}\TeX}
\newcommand{\filename}{esapub}

\date{16 november 2007}

\abstract
{}
{We provide a guideline to interpret the UVCS emission lines
(in particular O VI and Si XII) during shock wave propagation
in the outer solar corona.}
{We use a numerical MHD model 
performing a set of simulations of shock waves generated in the corona and
from the result we compute the plasma emission for the O VI and Si XII including the effects of NEI.
We analyze the radiative and spectral properties of our model
with the support of a detailed radiation model including Doppler dimming 
and an analytical model for shocks, and, finally,
we synthesize the expected O VI $1032\AA$ line profile.}
{We explain several spectral features of the observations like
the absence of discontinuities in the O VI emission during the shock passage,
the brightening of Si XII emission and the width of the lines.
We use our model also to give very simple and general predictions for 
the strength of the line wings due to the ions shock heating
and on the line shape for Limb CMEs or Halo CMEs.}
{The emission coming from post-shock region in the solar corona roughly agrees with
the emission from a simple planar and adiabatic shock,
but the effect of thermal conduction and the magnetic field may be important depending on the event parameters.
Doppler dimming significantly influences the O VI emission while Si XII line brightens mainly because of the shock compression.
Significant shock heating is responsible for the wide and faint component
of the O VI line usually observed which may be taken as a shock signature in the solar corona.}


\maketitle

\input{introduction.tex}
\input{model.tex}
\input{result.tex}
\input{conclusions.tex}
\input{acknowledgment.tex}
\appendix
\input{appendix1.tex}
\input{appendix2.tex}
\bibliographystyle{aa}
\bibliography{ref}

\end{document}

%% file: introduction.tex
\section{Introduction}\label{introduction}
Strong shock waves are expected to develop from fast Coronal Mass Ejections (CME).
According to the most popular idea about the structure of CMEs,
an expanding magnetic flux rope propagates in the solar corona and
this leads to the creation of the CME leading edge and to the creation of shocks, as well.
It is widely accepted that the front side halo CMEs are strongly correlated with geomagnetic disturbances
\citep{Brueckner1998,Cane2000,Gopalswamy2000b,Webb2000,Zhang2003}.
Thus, the detection of CME related shocks is a crucial topic in space weather research.
However, it is also a difficult problem,
because it cannot be confirmed from white light images alone.
A further problem is to describe the shock and figure out whether the environment
or the CME core expansion determine the shock wave properties,
since they influence the
acceleration of energetic particles as the shock evolves,
which is again a key question in space weather research.

All proposed hypotheses agree that the shock wave, whenever exists,
stays on the leading edge of the CME \citep{Hundhausen1987,Wagner1983}
though it may be too faint to be easily seen.
The leading edge, as observed in white light images, could be either plasma compressed by a shock
or denser material in magnetic loops ejected from near the solar surface.
Several shocks have been detected during CME front propagation with different diagnostics \citep{Ciaravella2006}.
Sharp edges in white light images have been taken as evidence of shocks at the CME leading edge,
but it is impossible to prove that they are shocks without the support of deeper diagnostics.
For instance \citet{Vourlidas2003} support the identification of shocks in SoHO/LASCO observations
thanks to an MHD model where a shock was needed to fit the observational constraints.
Sometimes the shock is detected without uncertainness by the presence of a Type II radio burst
that gives a reliable estimation of the post-shock density, but not of speed and location.
However, spectral analysis is the most general way to detect and describe the shock in the solar corona.

The UltraViolet Coronograph Spectrometer (UVCS) \citep{Kohl1995}
on board the Solar and Heliospheric Observatory (SoHO) \citep{Domingo1988}
gave a strong impulse to this research topic.
\citet{Ciaravella2006} pointed out the correlation between the shocks and several spectral properties of the post-shock emission.
Overall, they found that the shock emission is comparable or weaker than the background coronal emission.
The O VI 1032 $\AA$ line is usually broad, but
its intensity does not increase.
Instead the Si XII brightens significantly and it is common that the intensity of the line doubles,
and the Ly $\alpha$ fades.
This general description is based on several studies made on different CME events which reached similar conclusion.
For instance, broad O VI lines and more intense Si XII line have been observed in
2000 June  28 \citep{Ciaravella2005}, 2000 September 12 \citep{Suleiman2005} and
2000 October 24 \citep{Ciaravella2006} event.
In the shock related to the CME event on June 11, 1998 the derived compression and ionization
appear modest with respect to the inferred shock strength \citep{Raymond2000}.
Signatures of shock have been observed during the 2000 March 3 CME event \citep{Mancuso2002}.
They pointed out that the Oxygen temperature ($T_O$) was much
larger than the proton temperature ($T_p$) and that
the coronal O VI emission during a shock is well fitted by a narrow component which comes
from the quiet corona and a broad component which comes from the shocked plasma.
Actually, particles of different species can be heated by different amounts in a collision-less shock,
since there is not enough time for Coulomb collisions to bring the temperatures into equilibrium.
Observations of shocks in the solar wind generally show that electrons are weakly heated \citep{Schwartz2000}
and that minor ions of mass $m_i$ have higher temperatures than protons, perhaps as high as $T_i=(m_i/m_p)T_p$ \citep{Korreck2007}.
Observations of supernova remnants instead show $T_e \sim T_i$ for relatively slow shocks and $T_e \ll T_i$
for fast shocks \citep{Ghavamian2001} and
$T_O \sim T_p$ for slow shocks and $T_O \sim 8 T_p$ for a fast one \citep{Raymond2003a,Korreck2004}.\\
Since the whole sample of events amounts to $\sim 10$ shocks linked to CMEs
and only few of them have good enough data to do detailed analysis,
questions about the line emission from shocked plasma are still open.
Overall it is important to link all the observed features listed above with models, at least qualitatively.
Because of the large number of parameters that could characterize the evolution of a CME it is quite impossible
and even useless to model quantitatively one single event.
On the other hand, many recent efforts have focused on the theoretical modeling of the solar corona,
with the goal to properly describe the processes that lead to eruption and activity in the corona.
We think that modeling should be linked as much as possible to observables,
because several assumptions lay between the physics that the model describes and the actual emission
that could be inferred from that model. As an immediate implication, any 
attempt to model observables in shock wave propagation cannot neglect that the highly dynamic plasma 
can be easily in Non Equilibrium of Ionization (NEI) \citep{Spadaro1994}, since the time scales are too short to 
allow the ionization state of the plasma to relax.\\
The aim of this work is to find unambiguous spectral shock signatures, and
to provide a guide for the interpretation of general features that could appear in UVCS observations.
Our approach is to model in detail and diagnose the propagation of a shock wave generated
from a supersonic fragment of a CME in the magnetized corona.
We use a numerical Magnetohydrodynamic (MHD) model \citep{Pagano2007}
to describe the MHD evolution and 
from the result we compute the plasma emission for the O VI and Si XII lines which
are diagnostically important in UVCS observations of shock waves including the effects of NEI.
As a guideline to interpret the results, we support the study with a simple analytical model of a 
Rankine-Hugoniot planar and adiabatic shock wave.
From the radiation model we synthesize observable quantities, such as the profiles of spectral lines.\\
In Section \ref{model} we describe our model,
Sec.\ref{result} includes the results discussed
in Sec.\ref{conclusions}.

%% file: model.tex
\section{The model}\label{model}

We use a full MHD model of the solar corona tuned to investigate the coronal emission visible from UVCS
during a shock wave propagation.
We model a supersonic CME core fragment moving upward in a magnetohydrostatic solar corona. During its
propagation, the fast cloud generates a series of shock waves, and we study how the waves perturb 
the quiet corona. To this aim, we need to
model the evolution of both the plasma and the magnetic field and, to compute emission, also of the ion abundances. 

We consider the solar gravity, the radiative losses (e.g. \citet{Raymond1977},
a phenomenological coronal heating term and the field-oriented thermal conduction \citep{Spitzer1962}
as important physical effects. The full MHD equation we solve are:

\begin{equation}
\label{mass}
\frac{\partial\rho}{\partial t}+\vec{\nabla}\cdot(\rho\vec{v})=0
\end{equation}
\begin{equation}
\frac{\partial\rho\vec{v}}{\partial t}+\vec{\nabla}\cdot(\rho\vec{v}\vec{v})=-\nabla p+\frac{(\vec{\nabla}\times\vec{B})\times\vec{B}}{4\pi}+\rho\vec{g}
\end{equation}
\begin{equation}
\frac{\partial u}{\partial t}+\vec{\nabla}\cdot[(u+p)\vec{v}]=\rho\vec{g}\cdot\vec{v}-n^2P(T)+H_0-\vec{\nabla}\cdot\vec{F_c}
\end{equation}
\begin{equation}
\frac{\partial\vec{B}}{\partial t}=\vec{\nabla}\times(\vec{v}\times\vec{B})
\end{equation}

\begin{equation}
u=\frac{1}{2}\rho v^2+E
\end{equation}
\begin{equation}
\label{stato}
p=(\gamma-1)E
\end{equation}
with the constraint given by:

\begin{equation}
\label{divb}
\nabla\cdot\vec{B}=0
\end{equation}
where $t$ is the time, $\rho$ is the density, $n$ the number density, $p$ the thermal pressure,
$T$ the temperature ($T_e=T_p=T_i$ in this model), $\vec{v}$ the plasma flow speed,
$u$ the total energy (internal energy $E$ plus kinetic), $\vec{g}$ the gravity acceleration,
$\vec{F_c}$ is the conductive flux according to \citet{Spitzer1962}
and corrected for the saturation effect \citep{Cowie1977},
$P(T)$ the radiative losses per unit emission measure \citep{Raymond1977},
$H_0=n^2_0P(T_0)$ is a constant heating term whose only role is to keep steady the unperturbed corona
by balancing exactly its radiative losses,
with $n_0(\vec{r})=n(\vec{r},t=0)$ and $T_0=T(t=0)$ \citep{Pagano2007}.
We do not include plasma resistivity effects, which can be considered globally negligible on large scales in the solar corona.

We solve numerically the set of the ideal full MHD equations with the MHD module of the advanced parallel FLASH code,
basically developed by the ASC / Alliance Center for Astrophysical Thermonuclear Flashes in Chicago \citep{Fryxell2000},
with Adaptive Mesh Refinement (PARAMESH, \citealt{MacNeice2000}).
We include the FLASH module for the anisotropic thermal conduction \citep{Spitzer1962} implemented by \citet{Pagano2007},
and improved for the saturation effects by \citet{Orlando2005}

We compute the ionization state of the plasma from the history and distribution of $\vec{v}$, $n_e$ 
and T obtained with the MHD model. This is done asynchronously of the MHD computations;
we can do this safely because
we sample the solution on time bins ($\sim25$ s) shorter than a typical UVCS exposure time ($\sim120$ s) and
than the shock passage time scales ($\sim500$ s for a scale length of $5\times 10^9$ cm and a typical sound speed of $10^7$ cm/s)
and because we do not expect effects from small scale mixing by turbulent motion
in the presence of the magnetic field and of the thermal conduction.
The ionization state is computed considering the lagrangian transport of ions, and the ionization/recombination processes,
i.e. we are not assuming ionization equilibrium during the simulation.
The variation of the ionization fraction of each ion species is governed by the equation:

\begin{equation}
\label{ioniz}
\frac{\partial n_i^Z}{\partial t}=-\nabla\cdot n_i^Z \vec{v}+n_e[n_{i+1}^Z\alpha_{i+1}^Z+n_{i-1}^Zq_{i-1}^Z-n_i^Z(\alpha_i^Z+q_i^Z)]
\end{equation}

where $n_i^Z$ is the density of the element Z in the i-th ionization level in units of $cm^{-3}$,
$n_e$ is the electron number density and $\alpha_i^Z$ and $q_i^Z$ are respectively the recombination and ionization rates
for the element Z in the i-th ionization level in units of $cm^3/s$.

The ionization state evolution explicitly depends on electron and ion densities and the bulk velocity of the plasma,
and implicitly on the temperature through $\alpha_i^Z$ and $q_i^Z$.
In Eq. (\ref{ioniz}) the first term on the right hand side is the advection term,
and the second is the variation due to ionization and recombination processes that take place at the temperature
of the plasma element.
We solve this equation numerically for each relevant species and ions, assuming initial ionization equilibrium, 
and treating explicitly the advection
term with a Godunov scheme \citep{Godunov1959} and implicitly the second term with ionization rates given by \citet{Cox1985}
and recombination ones by \citet{Bryans2006}.
The evolution of the electron density and of the plasma motion throughout the domain are taken from the MHD model results
(sampled every 25 s).
A Godunov scheme is applied to compute the ion transfer through the cells due to the bulk motion of the plasma,
and the ionization fraction is updated considering the ionizing and recombining collisions (second term in Eq. (\ref{ioniz}))
between electrons and ions.

The intensity of the lines is computed as the sum of the collisional ($I_c$) and radiative ($I_r$) contributions 
(see the Appendix \ref{appendix1} for more details):
\begin{equation}
\label{intline}
I=I_c(n_e,n_i^Z,T) + I_r(n_i^Z,T,\vec{v})
\end{equation}

%
%
%
%
%

\begin{figure}[!htcb]
\centering
\includegraphics[scale=0.35,clip,viewport=20 10 450 600]{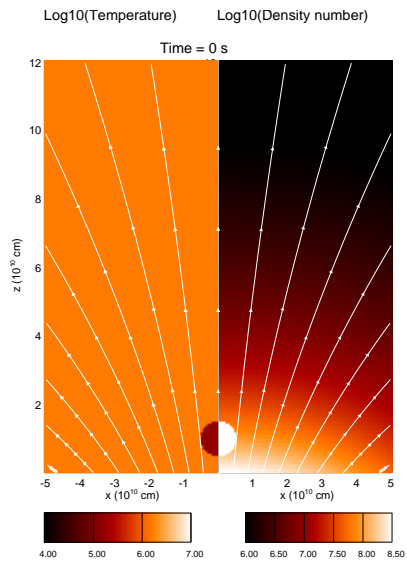}
\caption{Initial condition of the model.
The panel shows a section at $y=0$ of the temperature and density number spatial distributions.
The the magnetic field lines are also shown (white lines).}
\label{inicond}
\end{figure}

We model the evolution of a dense cloud moving upward supersonically in the outer coronal atmosphere as shown in Fig. \ref{inicond}.
The center of the cloud (i.e. the CME fragment) is initially positioned at $0.15 R_{\odot}$ above the photosphere and 
has an upward initial velocity.
It is denser than the surroundings, but in pressure equilibrium with them (i.e. colder than the surroundings atmosphere).
The atmosphere is isothermal at 1.5 MK and stratified by gravity in such a way to have $n=10^8 cm^{-3}$ at the initial CME 
fragment center height.
The background magnetic field is modeled as a dipole laying in the center of the Sun and seen 
from the central axis.
On this basic configuration, we build various setups of the other parameter and initial conditions.
As reference set up, the CME fragment has an initial velocity $v_c = 1000$ 
km/s and is 10 times denser than the surroundings (i.e. $n_c/n_0 = 10$ and
temperature: $T_c = 1.5\times10^5$ K).
The intensity of the dipole is chosen to give $\beta\sim 1$ at the initial height of the CME fragment center.
The other simulations that we present differ only for one parameter, and in particular, $\beta=10$,
$\beta=0.1$, $n_c/n_0=4$, $v_c=700$ km/s

\begin{table*}
\caption{Numerical simulations}
\label{tab:sim}
\renewcommand{\arraystretch}{0.8}
\begin{tabular}{l c c c}
\hline
\hline
Name & $\beta$ & $n_c/n_0$ & $v_c$ (km/s) \\
\hline
REF        & 1   & 10 & 1000 \\
$\beta 01$ & 0.1 & 10 & 1000 \\
$\beta 10$ & 10  & 10 & 1000 \\
$n_c4$     & 1   &  4 & 1000 \\
$v700$     & 1   & 10 &  700 \\
\\
\hline
\end{tabular}
\end{table*}

Simulations $\beta 01$ and $\beta 10$ are devoted to investigate the effect of the magnetic field on the global structure of the shocks, 
i.e. whether a stronger (weaker) magnetic field could lead to less (more) compression and, ultimately, to less (more) emission.
Simulation $n_c4$ and $v700$ explore situations in which the initial shocking fragment carries less momentum
(less mass the former, less velocity the latter).

The simulations are all performed in a 3-D cartesian domain ($x$,$y$,$z$)
that is $(8\times 8\times 16)\times10^{10}$ cm large,
for the duration needed by the shock to propagate well above $2 R_{\odot}$
which is the height where we put a hypothetical UVCS observations.
The time is $\sim1000 s$ when the initial velocity of the fragment is $1000$ km/s and $\sim1500 s$ when it is $700$ km/s.

%% file: result.tex
\section{Results}\label{result}

We now discuss our reference simulation and later the others which differ from it by just one parameter of
the initial conditions (i.e.: $\beta$, initial velocity of the cloud, density of the cloud).

\subsection{The reference simulation}
\begin{figure}[!htcb]
\centering
\includegraphics[scale=0.35,clip,viewport=20 10 450 600]{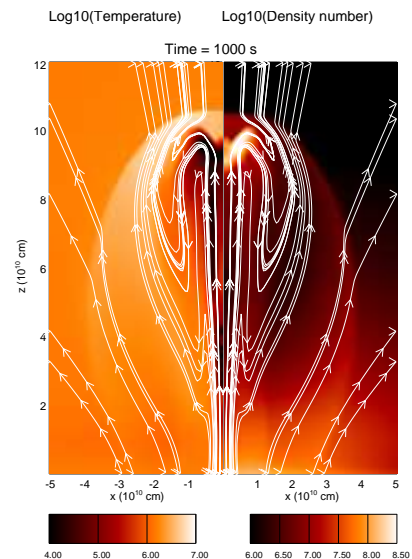}
\caption{As in Fig. \ref{inicond}, but at $t=1000$ s in the reference simulation.}
\label{simulationdens}
\end{figure}

In Fig. \ref{simulationdens} the density and the temperature of the plasma after 1000 s
of evolution are shown (for the reference simulation).
In all the simulations shock fronts depart from the upper part of the high speed cloud.
Here we address the shock evolution only, and not the cloud evolution.
The cloud continuously shocks the surrounding medium, since its speed remains faster than the sound speed
(at least during the time of our simulation), it
cools down during the motion because of the radiative losses and its core becomes thermally unstable.

\subsubsection{The Shock}
In the reference simulation, the shock propagates radially from the cloud.
Several shock fronts are generated 
during the passage of the cloud and finally they merge in a bow shock edge expanding in time.
After 1000 s of evolution the cloud has reached the height of  $\sim2.5 R_\odot$ and the produced shock edge has
traveled $\sim0.4 R_\odot$ from the originating cloud.
Behind the edge of the shock wave, the structure of the shocked gas is due to
the interaction and superposition of the different shock waves.

Since this work addresses the signatures of a shock transition during a CME, we focus our attention on 
the shocking edge which drives the disturbance of the quiet
outer corona rather than on the inner structure of the shock.
The closer in x direction the shocked plasma is to the shocking cloud,
the less the shock has propagated in the corona,
the stronger is the shock,
and the more the bulk velocity of the post-shock plasma is perpendicular to the solar surface.
The cloud velocity remains superalfv\'{e}nic and the magnetic field is shocked as well.
For this reason the post-shock magnetic field tends to be parallel to the shock front.
The plasma $\beta$ is relatively high in the upper corona ($\sim 15$ in our simulation at $2 R_{\odot}$),
and so the magnetic field pressure is negligible with respect to the plasma compression,
while the magnetic field orientation influences the thermal properties of the shock and, as consequence, the ionization state.
Since the shocked magnetic field is parallel to the shock front which is hotter than the surrounding atmosphere,
the thermal conduction toward pre-shock regions is ineffective and the shocked plasma does not cool down.
The shock heats the plasma, the ionization is enhanced and the ionization state changes.
Immediately behind the shock the plasma is far from the ionization equilibrium at the shock temperature,
and the ionization equilibrium is approached farther from the shock front,
where the temperature has been no longer changing much and the plasma is slowly relaxing.

\begin{figure}[!htcb]
\centering
\includegraphics[scale=0.35,clip,viewport=20 350 680 680]{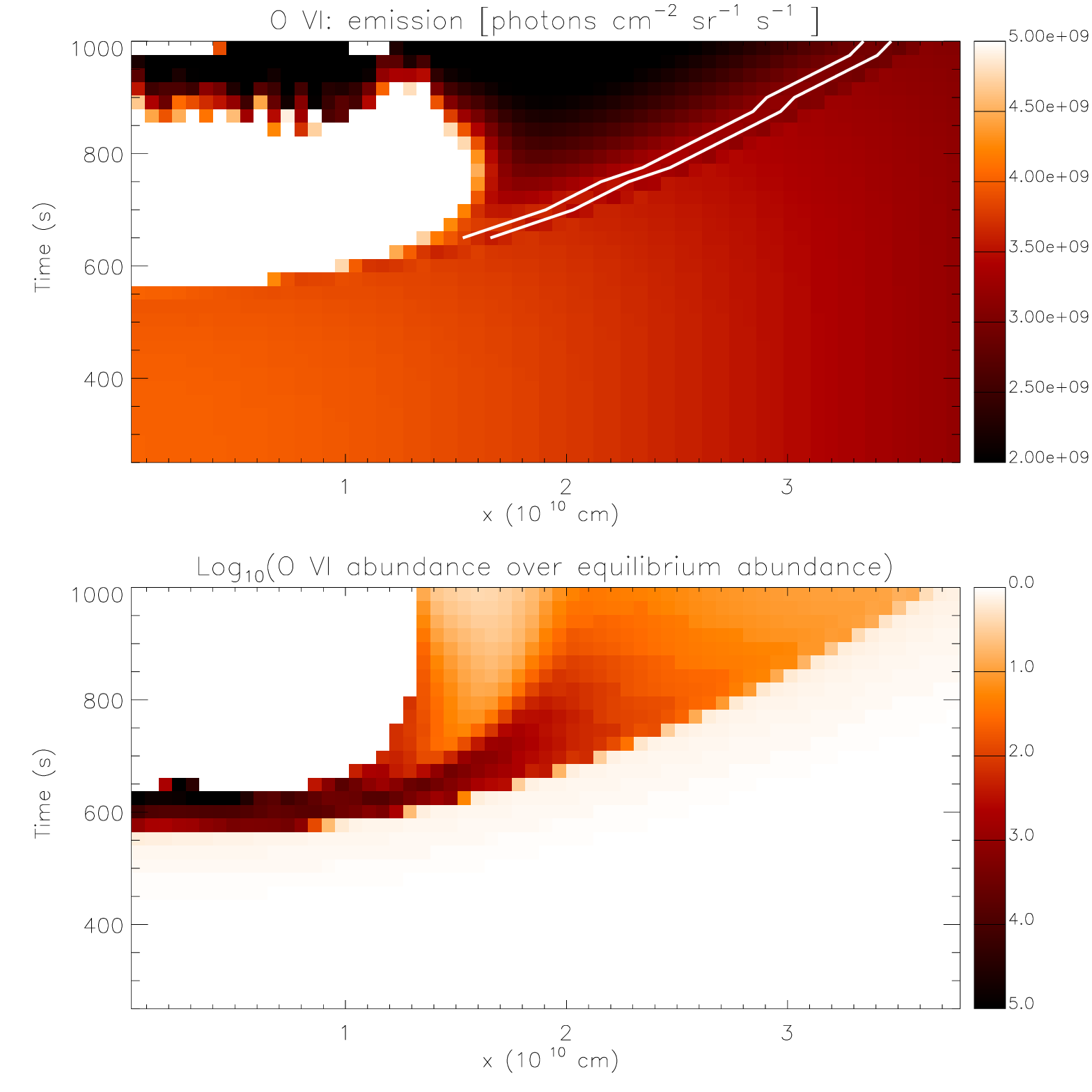}
\includegraphics[scale=0.35,clip,viewport=20 350 680 680]{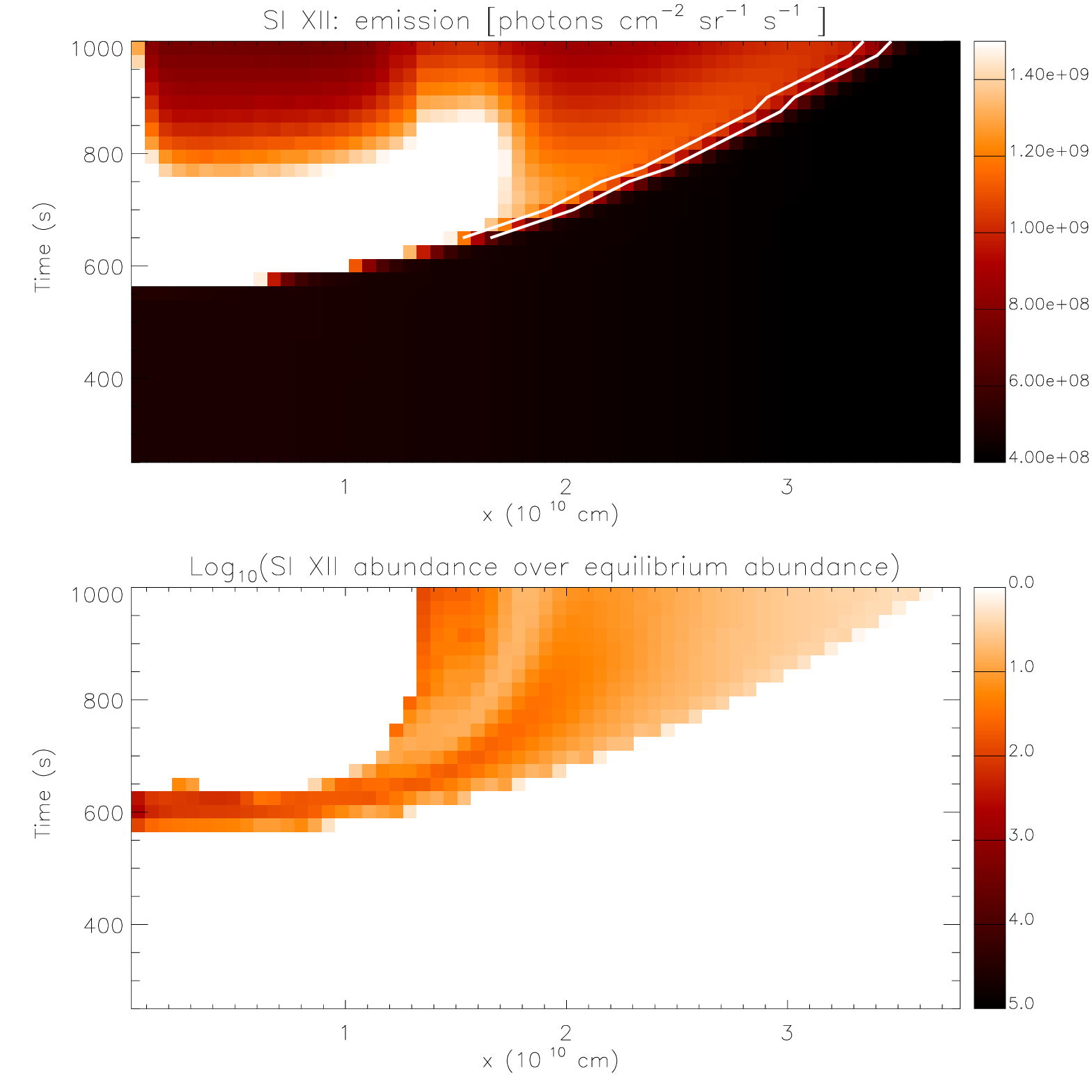}
\caption{Evolution of the emission of O VI (upper panel) and Si XII (lower panel) expected along an UVCS slit
positioned at $2 R_{\odot}$ integrated along the line of sight.
The evolution is sampled every 25 s.
The white lines mark the shock edge.}
\label{emission}
\end{figure}

\subsubsection{O VI and Si XII shock emission}
We apply our general radiation model to O VI and Si XII lines, that are important for the CME shock front diagnostics.
Fig. \ref{emission} shows the evolution of the line intensity across the UVCS slit computed
from the full MHD simulation and considering non-equilibrium of ionization.
The intensity jump across the shock is sharper in the Si XII line than in the O VI line.
Our model shows that the shocked plasma is far from ionization equilibrium; in fact,
the heating time due to the shock is much shorter than the ionization equilibrium time and there is not enough time for
the ionization state to change as soon as the shock has reached the plasma.
For instance, the ionization time for the O VI at the post-shock temperature ($T\sim3$ MK) is roughly $\tau_{eq}\sim 400$ s
(see Appendix \ref{appendix2}), significantly longer than the heating time.
We can estimate the actual heating time ($\tau_{heat}$) from the speed of the shock and the gyroradius of the particles.
Considering a magnetic field of $B\sim0.5$ G, a temperature of $\sim 2$ MK and the velocity of the shock ($\sim 500 km/s$)
we get $\tau_{heat}\sim 10^{-5}$ s.
Since the temperature in the region behind the shock is higher than the peak temperature for the O VI,
the emission becomes fainter and fainter behind the shock while the ions approach equilibrium,
i.e. at later time.
The shock heating makes the Si XII abundance increase, because the Si XII peak temperature is
higher than the temperature of the quiet atmosphere (1.5 MK).
Because of this, a large fraction of the Si XII emission comes from the inner part of the shocked region.

As mentioned above,
for diagnostic purposes, we focus the analysis at a characteristic height for the UVCS observations
and for shock formation \citep{Raymond2003b}, i.e. $2 R_{\odot}$.
It should be noted, that, in spite of the formation of the shock, the CME may either accelerate or decelerate
that far from the solar surface and that long after the ignition phase.
In order to investigate the role of the non-equilibrium ionization state, of the thermal conduction, and of the magnetic field,
we compare the intensity of the plasma behind the shock obtained from the MHD simulation with the one predicted 
by a simple model of
adiabatic and locally planar shock according to the Rankine-Hugoniot jump relations of density,
temperature and velocity as a function of the post-shock velocity of the plasma.
%
%
For further details about this analytical model see the Appendix \ref{appendix2}.
This simple model is a rough approximation for our shocks, since the shock is not propagating in a very low beta plasma
(neglect the momentum flux due to the Lorentz force), the thermal conduction is mostly inhibited by the magnetic field 
(adiabatic plasma)
and the shock heating is much faster than the plasma ionization and recombination
(the ionization fraction does not change at the shock front).
Significant departures from this approximation may be signature that some of the assumptions listed above do not hold.

\begin{figure}[!htcb]
\centering
\includegraphics[scale=0.25,clip,viewport=0 0 1000 700]{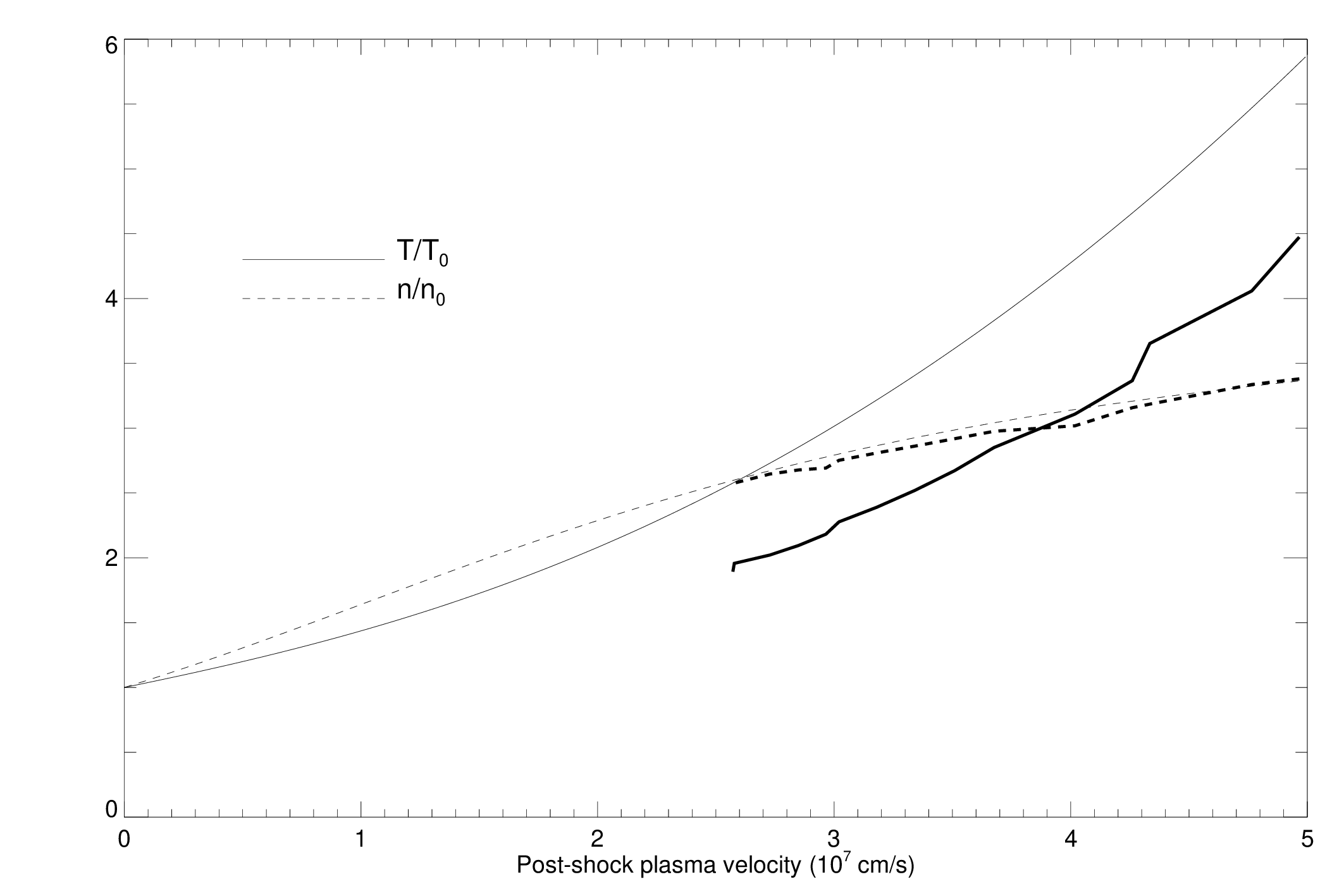}
\caption{Ratio between post-shock and pre-shock temperature (solid lines) and density (dashed)
as a function of the post-shock plasma velocity. 
Thick lines mark results obtained from the MHD simulation, and thin lines from an analytical model
of Rankine-Hugoniot relations} on an adiabatic and planar shock.
\label{shockrankhugnT}
\end{figure}

To compare the detailed simulation with the analytical model so as to evaluate the deviations, 
in Fig. \ref{shockrankhugnT} we plot the ratios between the post-shocked and pre-shocked density and temperature
as function of the post-shock velocity of the plasma for the MHD simulation and for the analytical model.
The post-shock quantities for the simulation (i.e. post-shock velocity, density and temperature jump)
are computed at the shock edge, marked by white lines in Fig. \ref{emission}.

The front of the bow shock begins to cross the UVCS slit field of view at $\sim675$ s.
UVCS detects first the upper and fastest part of this front and later the flanks.
As mentioned above, the shock strength decreases in time during the crossing of the UVCS slit field of view and
at t=700 s the plasma is accelerated to $\sim500$ km/s because of the shock and
to $\sim250$ km/s at $t=1000$ s (see Fig. \ref{shockrankhugnT}).
As shown in Fig. \ref{shockrankhugnT} the density jump in the simulation roughly agrees with that of the analytical model,
meaning that the magnetic field has little relevance in this dynamic regime.
The temperatures agree with less accuracy because the thermal conduction is not completely ineffective,
and the shock can cool down for that.
For instance, the thermal conduction is effective in the region where the shock is stronger,
i.e. where the shock propagation is not perpendicular to the post-shock magnetic field.
Here (i.e. $x\sim 2\times 10^{10}$ s and $z\sim 10\times 10^{10} $ cm in Fig. \ref{simulationdens}),
the angle between the magnetic field and the thermal gradient is $\sim30^{\circ}$.
Considering the thermal gradient, the temperature and the length scale involved, we estimate a thermal conduction
time scale of $\sim5$ s to be compared to the
dynamical (expansion) time on the same length scale which is $\sim100 s$.
The thermal conduction is therefore very effective and leads to temperatures lower than those predicted by
the Rankine-Hugoniot relations.

\begin{figure}[!htcb]
\centering
\includegraphics[scale=0.25,clip,viewport=0 0 1000 700]{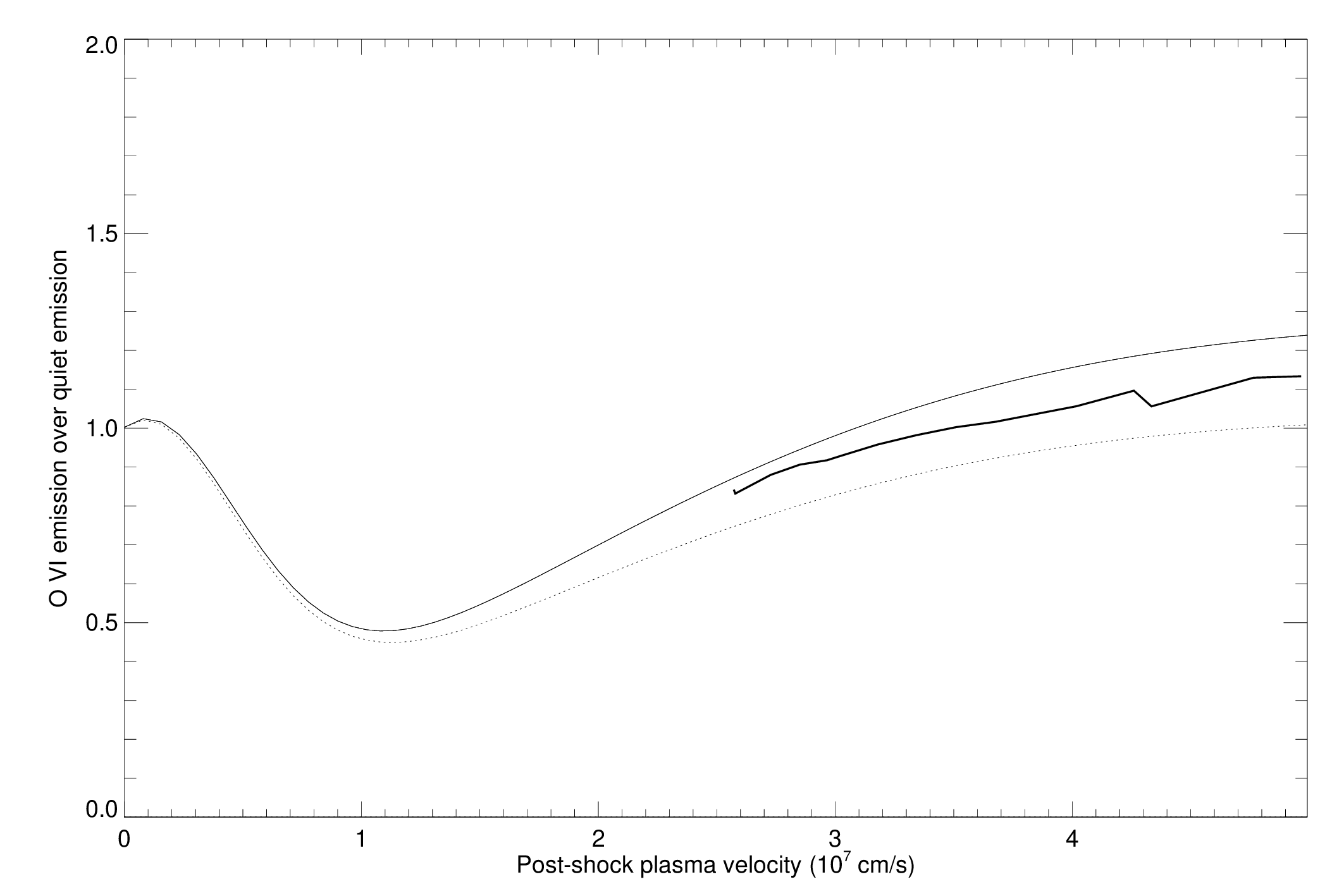}
\includegraphics[scale=0.25,clip,viewport=0 0 1000 700]{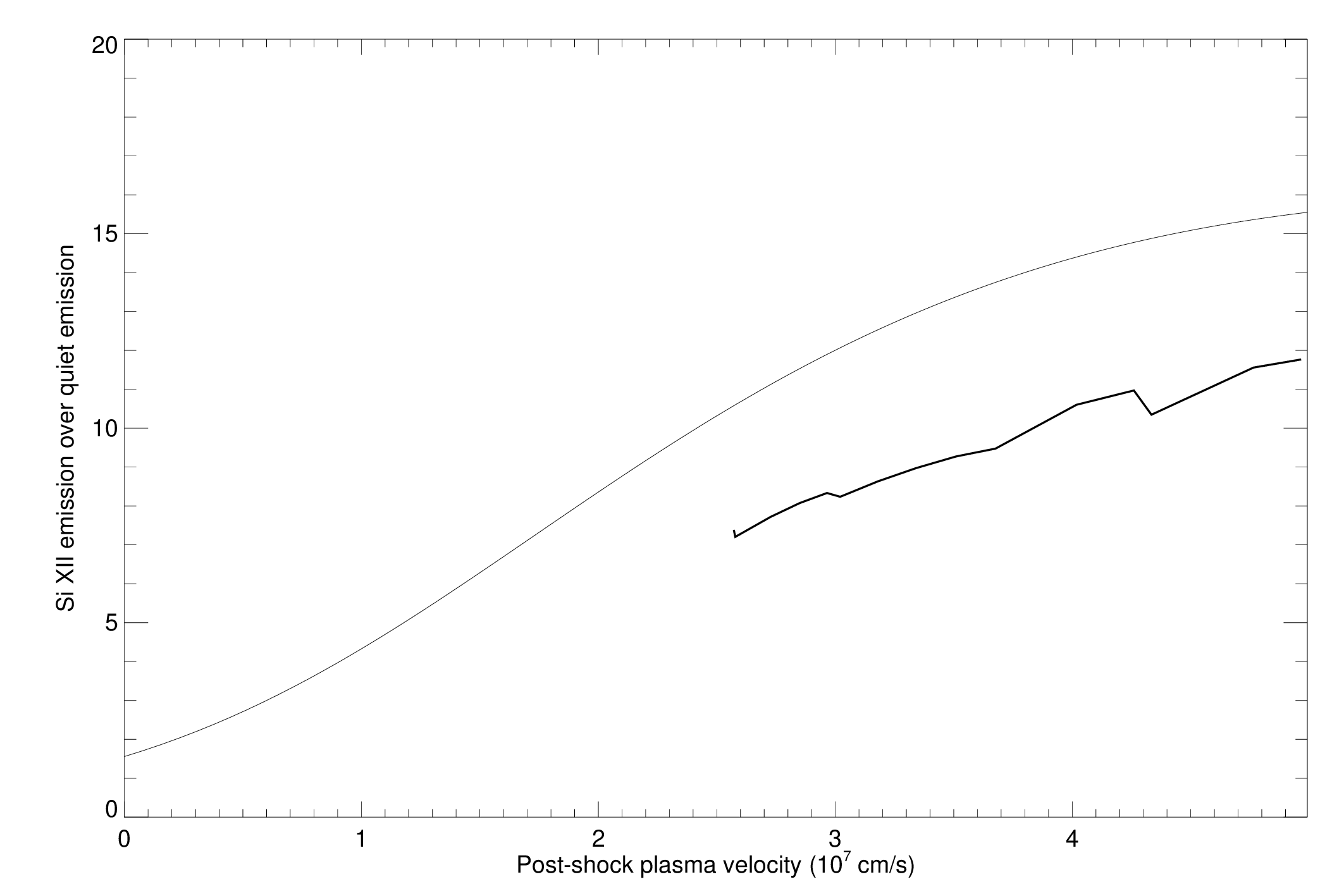}
\caption{Ratio between post-shock and pre-shock emission in the O VI (upper panel) and Si XII (lower panel) lines
as a function of the post-shock plasma velocity.
Results from the MHD simulation (thick lines) and from the analytical model (thin lines) are shown.
In the latter case we consider approximated ionization state formulae
assuming $\tau=0$ (solid line) or $\tau=100$ (dashed line) and our radiation model (see Appendix \ref{appendix1}).}
\label{shockrankhugI}
\end{figure}

From the jump in the density and the temperature we derive the jump in the emission for O VI and Si XII 
with our radiation model.
It is plotted in Fig. \ref{shockrankhugI} as a function of the post-shock plasma velocity
(O VI upper panel and Si XII lower panel).
In the O VI line the radiative part of the line is significant for the unperturbed plasma,
but it completely vanishes because of the Doppler dimming in the region behind the shock.
The radiative contribution to O VI 1032 is 4 times that of O VI 1037
(in the absence of other pumping), while the collisional component is a factor of 2 brighter.
The shock compression makes the collisional contribution of the line increase,
whereas the shock acceleration makes the radiative contribution decrease at the same time.
For this reason, the shock shows an emission fainter than, or at most comparable with, 
the quiet corona, as usually observed \citep{Ciaravella2005,Ciaravella2006},
i.e. O VI emission ratio $\sim1$ in Fig. \ref{shockrankhugI}.
The weight of the radiative part depends on the non-shocked plasma density and in our model
considering an unperturbed density of $\sim2\times10^6 cm^{-3}$, it is 85\% of the emission
and increases as the unperturbed density decreases.
Instead, since the radiative part of the Si XII is negligible,
the total intensity of the
post-shock line emission jumps significantly and relatively to the strength of the shock,
as plotted in Fig. \ref{shockrankhugI} (lower panel).
As a consequence, the
Si XII line clearly brightens,
while the O VI line does not, as \citet{Ciaravella2006} pointed out.

\subsection{Comparison with other simulations}
\subsubsection{Different $\beta$}
As anticipated above, we perform two more simulations considering different intensities of the magnetic field,
with a less ($\beta\sim 10$) and more ($\beta\sim 0.1$) intense magnetic field than in
the reference simulation, respectively.
\begin{figure}[!htcb]
\centering
\includegraphics[scale=0.25,clip,viewport=0 0 1000 700]{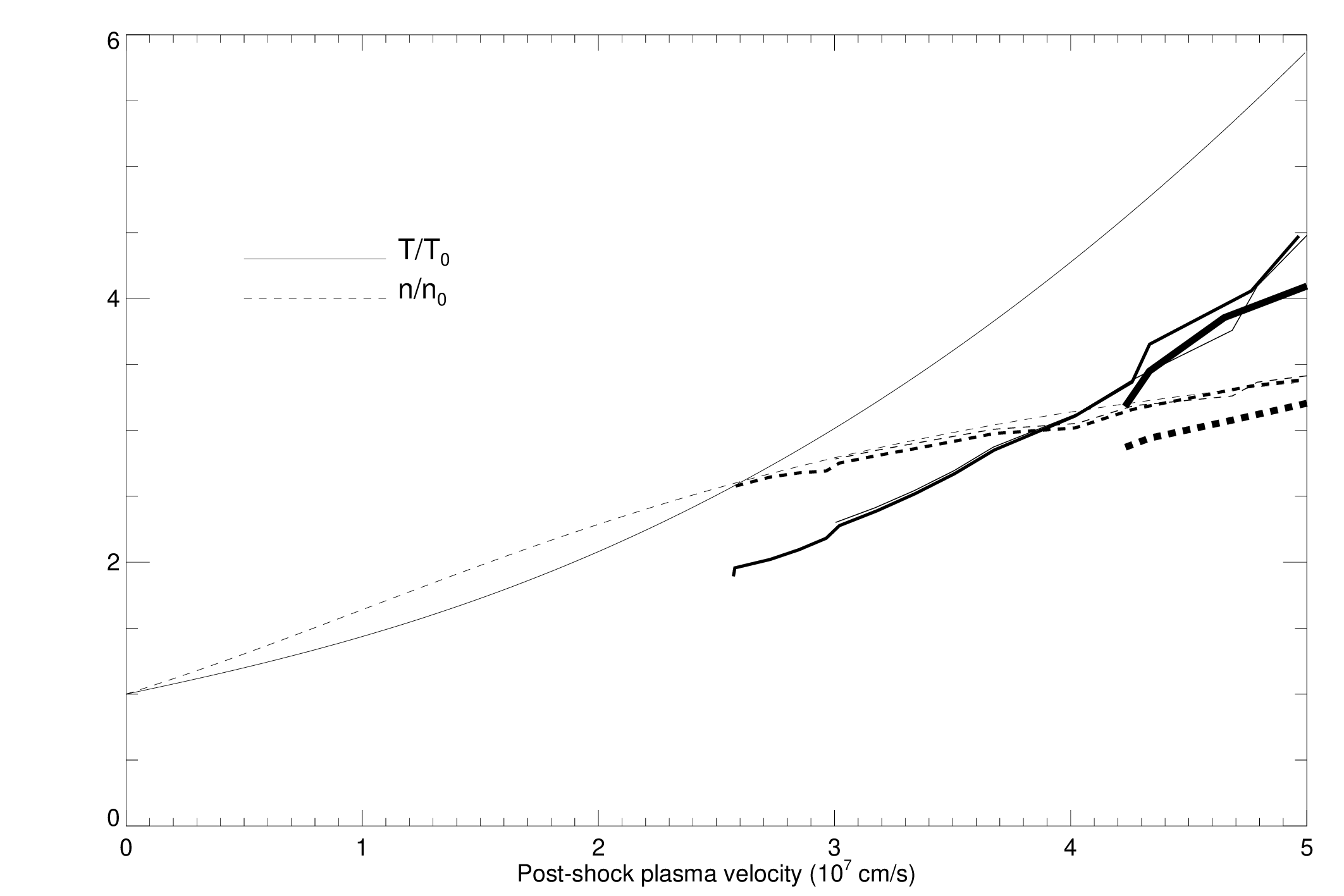}
\caption{Ratio between post-shock and pre-shock temperature (solid lines) and
density (dashed lines) as function of the post-shock plasma velocity.
The thickness of the line is inversely proportional to the initial $\beta$ of the simulation,
($\beta=0.1$, $\beta=1$ and $\beta=10$).
The thinnest lines is for the ratio computed analytically from Rankine-Hugoniot relations}.
\label{shockrankhugbetanT}
\end{figure}
Fig. \ref{shockrankhugbetanT} shows the density and temperature jump across the shock 
for the simulations with different values of $\beta$.
The simulation with $\beta=10$ shows a very similar behaviour
and the shock properties do not depart significantly from the reference case already discussed.
This is not surprising, since we have already shown that the magnetic field has
no relevance for the shock dynamics in this regime.

In a stronger magnetic field (Fig. \ref{shockrankhugbetanT}),
the shock is significantly influenced by the magnetic pressure.
According to the theory of MHD shocks the compression due to the shock
is splitted between magnetic field and plasma and, as result,
the plasma density jump is weaker than in a normal HD shock \citep{Bazer1959}.
This effect characterizes our simulation with a stronger field.
The density jump from the pre-shock to the post-shock region is $\sim5$\%
less than reference case,
but the temperature does not show any significant departure
since the energetics of the shock is not influenced by the magnetic field
except for the partial thermal insulation which acts regardless of the $\beta$ value.
The radiative properties of this case descend from these considerations.
The emission of the O VI and Si XII line is weaker than in the reference case, because the density is lower.
The radiative part of the line vanishes because of the Doppler dimming,
and the intensity of the line is weaker than the reference case is by $\sim10$\%.

\subsubsection{Weaker cloud}\label{weakercloud}
We now discuss the propagation of shocking clouds that initially carries less momentum,
i.e. with initial density contrast $n/n0=4$, instead of $n/n0=10$, or initial cloud velocity 
$v_0=700$ km/s, instead of $v_0=1000$ km/s.
The shock propagation is slower in those simulations and the top of the cloud needs
$600$ s and $875$ s to reach the height of $2R_{\odot}$, 
in the simulation with $n/n0=4$ and in the one with $v_0=700$ km/s respectively.
During the travel of the cloud in the corona, the shock departs from it radially
reaching a final distance from the cloud of $\sim 0.6 R_\odot$ in the x direction.
The final bow shape of the shock front is similar to the one in the reference case.
Of course, the smaller the energy of the cloud, the less the shock strength at any time,
so that the plasma is never accelerated over $\sim300$ km/s.
The velocity relative to the cloud is 30\% smaller and that leads to a corresponding delay 
in the detection at $2R_{\odot}$.
This delay allows the plasma to better approach ionization equilibrium.


\subsection{O VI line profile}
To further investigate the observational constraints for the shock
properties we show here the line profile of the O VI lines synthesized
from our reference model.
This line is analyzed in detail by \citet{Mancuso2002}
who indicate shock diagnostics from UVCS observations.
Here we try to provide further insight to that analysis,
but our synthesis of the line profile is very general and 
could be applied also to the Si XII line
after tuning the appropriate parameters.

\begin{figure}[!htcb]
\centering
\includegraphics[scale=0.55,clip,viewport=40 10 500 341]{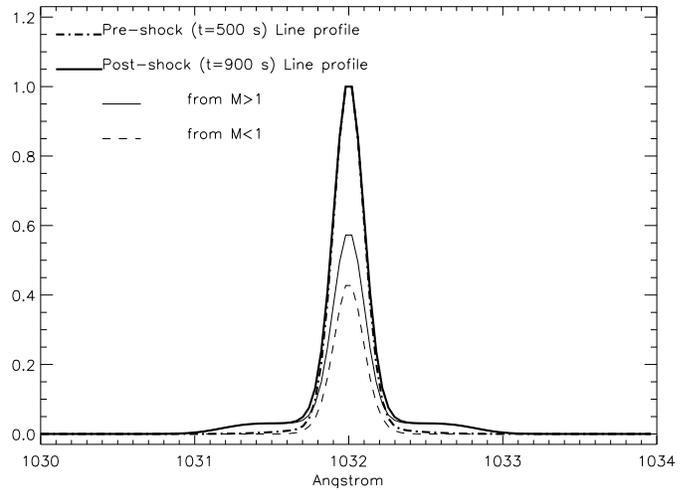}
\caption{Line profile of the O VI line synthesized from our reference model if observed as
a limb CME at $t\sim500$ s (point-dashed thick line) and at $t\sim900$ s (solid thick line),
i.e. before and after the shock crossed the UVCS field of view, respectively.
The emission is averaged on time bins of 150 s and on space bins of
9' along the UVCS slit through which the CME is observed.
The profile is normalized to the peak.
We also show the two components which combine to form the line at $t\sim900$:
the one emitted by the shocked plasma ($\mathcal{M}>1$, solid thin line)
and the one emitted by the unshocked plasma ($\mathcal{M}<1$, dashed thin line).}
\label{lineprofilelimb}
\end{figure}

We take into account the thermal broadening of the line and the Doppler
shift due to the plasma motion projected along the line of sight, but not
the instrumental broadening. To approach a realistic UVCS observation
we average the spectrum over time bins of 150 s and over space bins of 9'.
We also consider integrations along two different lines of sight: one
mimicking a limb CME, the other a halo CME.
Finally we discuss the effect of the differential Oxygen/protons shock heating on the Limb CME line profile.

\subsubsection{Limb CME}
In Fig. \ref{lineprofilelimb} we plot the profile of the O VI line for
the reference model as if it were observed as a limb CME before and after
the shock passed the UVCS slit.  
For the moment we assume equal proton, electron and ion temperature.
After the shock
the central part of the
line has not changed because it is emitted by that part of the plasma 
never involved in the shock passage. 
Evidence of the passage of the shock 
across the UVCS slit can instead
be found in the wings of the later line. They are more prominent than in the
quiet line, because of the line shift due to the line
of sight component of the shock velocity.
The resulting line profile is the sum of the emission contribution from un-shocked plasma and shocked plasma.
The un-shocked plasma emits only a narrow component (Fig. \ref{lineprofilelimb}, $\mathcal{M}<1$ line),
while superposition of the lines emitted by the shocked plasma gives both a central component and significant wings.
These two emission contributions merge in a two component line.
The first is a narrow and bright component, the sharp peak emitted by the unshocked corona
and by the shocked plasma that moves perpendicularly to the line of sight;
the second component is due to the shocked plasma that has a significant component of the velocity parallel to the line of sight,
so that it emits thermally broadened lines centered far from $1032\AA$.
The relative weight of the two
components is strongly sensitive to the background emission, which
seriously influence the narrow component. \citet{Mancuso2002}
indeed observed a similar line profile and concluded that a shock was
crossing the field of view of the UVCS slit during the observation.
We argue that a two component line profile, as shown in Fig.
\ref{lineprofilelimb}, may in general be a signature of the presence of
a shock.

\subsubsection{Oxygen shock heating}
\begin{figure}[!htcb]
\centering
\includegraphics[scale=0.55,clip,viewport=40 10 500 341]{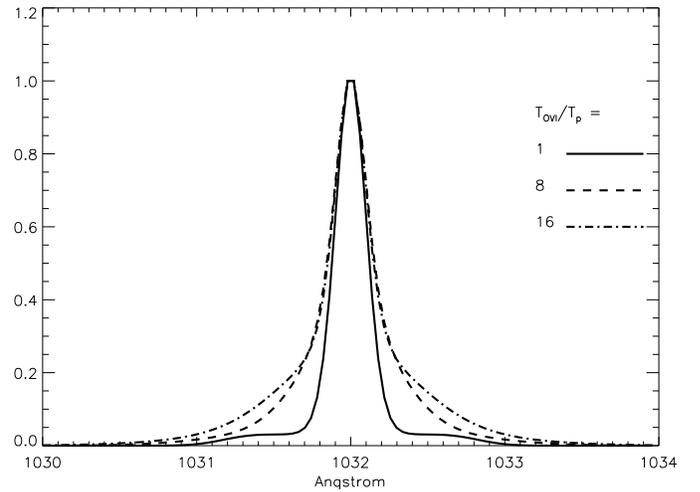}
\caption{As in Fig. \ref{lineprofilelimb}, but considering
different Oxygen-electron plasma temperature ratio, as indicated in the legend.
All the line profiles are computed at $t\sim900$s.}
\label{lineprofilelimbheat}
\end{figure}

We expect that the wings of the line profile become more prominent if the shock heating
mechanism is more effective on the heavy ions than on the protons,
because the thermal broadening becomes larger right in the Doppler-shifted components of the ion lines.
To test this consideration,
already made by \citet{Korreck2004} and \citet{Raymond2003a}, we check what
happens if we artificially increase
the O VI temperature obtained from our MHD model by a given factor.
Fig. \ref{lineprofilelimbheat} shows the O VI line profile,
obtained by increasing the temperature by a factor
8 \citep{Korreck2004,Raymond2003a} and 16 (oxygen atomic mass).
Because of the temperature decoupling between electrons and ions, the
FWHM of the line emitted by the shocked plasma increases from
$\sim 0.2\AA$ up to $\sim 1 \AA$.

\begin{figure}[!htcb]
\centering
\includegraphics[scale=0.51,clip,viewport=0 10 500 341]{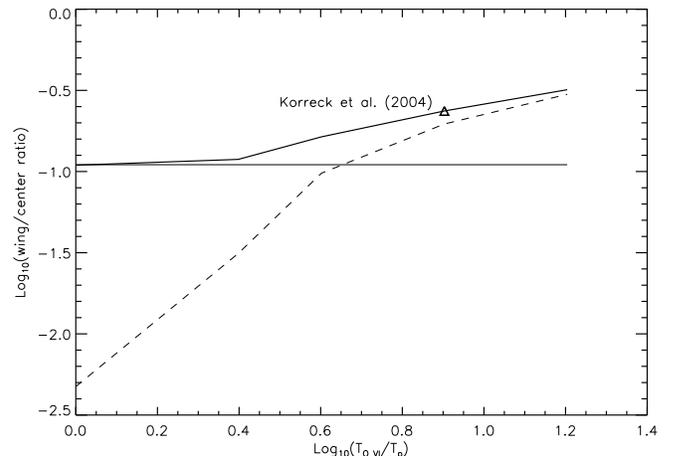}
\caption{Fraction of the O VI line emitted in the wings
as function of Oxygen-electron plasma temperature ratio.
The thin solid line is considering only the Doppler effect (temperature of the atmosphere not changed),
the dashed line only the thermal broadening, the thick solid line both of them.}
\label{wings}
\end{figure}

The wings of the line profile brighten if a greater $T_{OVI}/T_p$ ratio is supposed.
Fig. \ref{wings} shows the ratio between the total
emission of the wings (i.e. emission out of the range $1032\pm 0.3 \AA$)
and that of the center of the line (i.e. inside that range).
It increases from 0.1 ($T_{O \ VI}=T_e$) up to 0.24 ($T_{O \ VI}=16T_e$).
The trend is mostly due to the increasing thermal broadening of the emission 
from the post-shock plasma (Fig. \ref{wings}, dotted line), but it
is limited by the presence of the other broadening component 
which is independent of the temperature (Fig. \ref{wings}, thin solid line).
The sum of the two effects gives the total strength of the wings 
in the observed lines (Fig. \ref{wings}, thick solid line).
The stronger is the shock, the stronger are the wings.
Therefore at $\sim575$ s, when the first part of the shock crosses the UVCS slit field of view
and the shock is the strongest,
the wing/center ratio is $\sim0.6$ and later,
at $\sim1000$ s, it decreases down to 0.24 (shown in Fig. \ref{wings}).
It should be noted that Fig. \ref{lineprofilelimbheat}
shows that $T_O/T_p = 16$ resembles the profiles shown by \citet{Raymond2003a}
and \citet{Mancuso2002} far better than the lower temperature
ratios, supporting the assertions of those papers that $T_O/T_p$ was large.
Our approach, when applied to specific
events, can give
reliable and alternative measurements on the actual $T_i/T_p$,
once measured the post-shock electron temperature, and of the components of the velocity,
since the value of the wing/center ratio uniquely corresponds to a certain ion temperature.

\subsubsection{Halo CME}
\begin{figure}[!htcb]
\centering
\includegraphics[scale=0.55,clip,viewport=40 10 500 341]{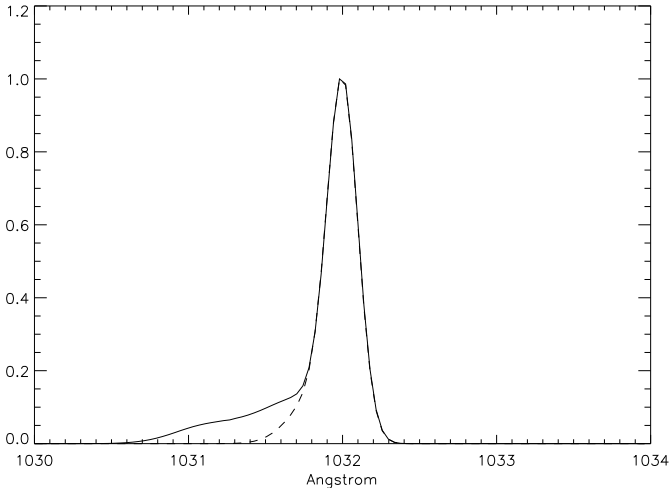}
\caption{Line profile of the O VI line synthesized from our reference model if observed as
an Halo CME at $t\sim900$ s, i.e. after the shock crossed the UVCS field of view.
The emission is averaged over time bins of 150 s and over space bins of
9' along the UVCS slit through which the CME is observed. The profile is normalized to the peak.
For the solid line the slit is $\sim 5'$ apart from the center of the Sun, $\sim 7'$ for the dashed one.}
\label{lineprofilehalo}
\end{figure}

Fig. \ref{lineprofilehalo} shows the results for the reference model
if oriented as a Halo CME expelled toward the observer,
not considering any temperature increase for the ions. In the line
profile there is only one tail on the blue side.
The line of sight is appropriate to detect the Doppler shift due to
the global front motion, rather than that due to front expansion,
detected for the Limb CME line of sight.
At increasing distance from the shock central axis the blue-shift is 
progressively reduced 
due to the projection effect.
The total line profile consists of the superposition of radiation with
continuously varying blue-shift.
Fig. \ref{lineprofilehalo} shows the line detected by the slit pointing 
at $\sim 5'$ (solid line) and $\sim 7'$ (dashed line) from the center of the Sun.
The closer the field of view to the Sun disk,
the larger is the line of sight component of the post-shock bulk velocity and
the more blue-shifted is the emission.
The line presents a very bright quiet emission and 
an envelope of shifted lines extending to $\sim 1 \AA$ from the quite line
(see Fig. \ref{lineprofilehalo}, solid line).
Although the details of the structure of the line are strongly related 
to the structure and geometry of the shock, the very presence of this
smooth and asymmetric component of the line could be a distinctive feature,
although such observations are
technically unfeasible for UVCS, because the slit is too close to the
center of the Sun, we may equally hope to detect this feature in
real observations.  In fact, since the shock front is more than 16' wide,
if the CME is not expelled perfectly along the Sun-Earth line,
part of the shock front could be visible out of the Sun
disk in a realistic UVCS field of view.
This conclusion would apply to the halo CME shocks observed by \citet{Ciaravella2006}.
Moreover, as we already claim above (Section \ref{weakercloud}), a
shock front could extend over a wider angle if the shocking fragment
is less dense.

Finally, for CMEs expelled in directions in between the
extreme cases considered here we expect to detect both the broadening
effect due to the thermal broadening of the shocked plasma (Limb CME)
and the one due to the Doppler shifted emission from fast plasma moving
toward the observer.


%% file: conclusions.tex
\section{Discussion and conclusions}\label{conclusions}
In this work we study the shock fronts departing from supersonic CME fragments with
detailed MHD modeling. We address the diagnostics obtainable with
UVCS observations, which have been widely analyzed in this framework,
and in particular the O VI and Si XII emission lines.
As a result, we are able to explain most of the spectral signatures of
the shocks revealed by the observations, and we step forward to predict
more shock properties. In particular we indicate the main differences in
the line shapes expected between halo CMEs and limb CMEs and the strength
of the wings of the O VI line as possible diagnostics for ion shock
heating.

We have chosen as a reference case the shock front originating from a CME
fragment 10 times denser than the surrounding 1.5 MK magnetized ($\beta
\approx 1$) corona expelled at a speed of 1000 km/s. The ambient magnetic
field is a large scale dipole. Then we have explored other cases with
higher and lower magnetic field, with less dense cloud and with slower cloud.
For each simulation we have computed the emission in the selected
lines and derived the line profiles with a detailed radiation model
including the radiative and collisional contributions and
the effects of non-equilibrium of ionization.

From the reference simulation we find that
some of the distinctive shock features detectable in O VI and Si
XII lines arise from the relative brightness of the collisional and
radiative contributions in the unperturbed corona.  Since, the radiative
contribution is actually negligible for Si XII, but important for O VI,
the intensity jump across the shock is sharper in the Si XII line than
in the O VI line. The intensity of the Si XII line scales roughly as the
square of the compression, while the one of the O VI scales more smoothly
because its emission is strongly biased by the radiative contribution.
In the O VI line the radiative part completely vanishes 
in the region behind the shock because of the Doppler dimming, 
while the shock
compression makes the collisional contribution of the line increase. The
two effects are competitive and comparable, so that the shock shows an
emission fainter than or comparable with the quiet corona in this line. 
In the Si
XII line, the shocks produce only the growth of the collisional
contribution. This explains why at the same time the Si XII line clearly
brightens, while the shock has little influence on the O VI line.
These considerations explain past observations.
\citet{Ciaravella2005} found that during the shock passage in the
2000 June 28 event the O VI emission was weaker or comparable to the
quiet corona emission. \citet{Ciaravella2006} examined several events
and showed that it occurs commonly that when a shock is detected the O VI does
not brighten, and the Si XII brightens significantly.

The ionization fractions of oxygen and silicon do not change in shock
transitions because the shock heating and passage are much faster that
the ionization/recombination processes. In our simulations the shocks
are not extremely strong, so that they represent realistic situations,
typically observed.
For this reason, we expect that the conditions of 
ionization non-equilibrium are commonly found and that
the ionization fractions are comparable to those of the quiet corona.
Behind the shock front the plasma should slowly
approach the ionization equilibrium at post-shock temperature.

The effect of the magnetic field on the shocks depends on the strength
of the field. In high $\beta$ plasma the role of the magnetic field
is only to thermally insulate the shocked plasma.  It happens because
in this regime the shock is superalfv\'{e}nic and it is responsible for
the reorientation of the magnetic field parallel to the shock front,
i.e. perpendicular to the thermal gradient. In low $\beta$ plasma,
the magnetic field influences also the compression of the plasma.
In this regime (also superalfv\'{e}nic) the compression due to the shock
is split between the magnetic field and the plasma; therefore, the
density (and the line intensity) is lower than in a high $\beta$ regime.
\citet{Raymond2000} argued that a relatively strong magnetic
field could in principle be responsible for modest compressions with
respect to the observed shock speed.

Differences in the kinetic energy of the CME could lead to different
extent of the shock front. We find that the weaker the CME core,
the more extended and slow is the shock propagation.

When we observe the shock propagating from a limb CME we expect to observe
a profile of the O VI line consisting on two distinct overlapping contributions.
The unperturbed plasma that lies along the line of sight emits a narrow
line and
the shocked plasma emits a wider line.
This combination is the kind of line shape
that \citet{Raymond2000} and \citet{Mancuso2002} observed during the CME events.
They used the double component line profile as signature of the shock.
In our simulation we get the same line profile shape after
the shock passage.  Therefore we confirm that this spectral feature is a
reliable signature of the shock presence.

In our simulation we have a line of sight component of the velocity
of $\sim200 km/s$, leading to a line shift of $\sim0.7$ \AA. The
shock heating $T_O=T_p$ leads to a line shift of $\sim0.2$ \AA.
The acceleration of the plasma is responsible for the presence of
the wings and higher $T_O/T_p$ leads to more significant wings.
For this reason, diagnostic of the ion shock heating could be based 
on measuring the relative
importance of the wings of the O VI line and on estimating the
post-shock electron temperature, and the components of the velocity.  
Here we present an example
of proton and oxygen enhanced heating
\citep{Korreck2004,Raymond2003a}.
It will be interesting in the future
to apply this kind of diagnostics of the ions shock heating to
one particular event.

Finally, we expect to see an asymmetric O VI line profile with a smooth
tail on the blue-side, when the shock propagating from a halo CME expelled
toward the observer enters in the UVCS field of view.
A condition for the detectability of this effect out of the solar disk are a large line of
sight component of the plasma velocity and a wide shock front.

%% file: acknowledgment.tex
\begin{acknowledgements}
The authors thank Angela Ciaravella for fruitful discussion and feedback on observational aspects.
They acknowledge support for this work from Agenzia Spaziale
Italiana (contract I/035/05/0),
Istituto Nazionale di Astrofisica and Ministero dell'Universit\`a e Ricerca.
P. Pagano's stay at the Center for Astrophysics
was supported by NASA grant NNG06GG78G to the Smithsonian Astrophysical
Observatory. 
The software used in this work was in part developed by the DOE-supported ASC / Alliance Center for Astrophysical Thermonuclear Flashes
at the University of Chicago, using modules for thermal conduction and optically thin radiation built at the Osservatorio Astronomico di Palermo.
The calculations were performed on the Exadron Linux cluster at the SCAN (Sistema di Calcolo per l'Astrofisica Numerica) facility of
the Osservatorio Astronomico di Palermo and on the IBM/SP5 machine at CINECA (Bologna, Italy).
Part of the simulations were performed within a key-project approved in the INAF/CINECA agreement 2006-2007.
CHIANTI is a collaborative project involving the NRL (USA), RAL (UK), MSSL (UK), the Universities of Florence (Italy) and Cambridge (UK), and George Mason University (USA).
\end{acknowledgements}

%% file: appendix1.tex
\section{The radiative and collisional contribution \label{appendix1}}
The intensity of the lines is modeled as the sum of the collisional ($I_c$) and radiative ($I_r$) contributions
as functions of $n_e$, $n_i^Z$, $T$ and $\vec{v}$:
\begin{equation}
\label{intlineapp}
I=I_c(n_e,n_i^Z,T) + I_r(n_i^Z,T,\vec{v})
\end{equation}

\subsection{Collisional contribution \label{collisional}}
The collisional contribution is:

\begin{equation}
\label{IC}
I_c(n_e,n_i^Z,T) = n_e n_i^Z Q(T)
\end{equation}

where $Q(T)$ is the collisional excitation rate coefficient for
the ions which treats the collisionally induced emission of the ions.
Here we present the approximated formula for Q(T) we use for the two lines of interest for this paper
(i.e. O VI $1032\AA$ and Si XII $499\AA$):

\begin{equation}
\label{Q(T)OVI}
Q(T)_{O \ VI}=Q_0^{O \ VI} \frac{e^{-T_{cut}^{O \ VI}/T}}{\sqrt{T}}\ln{\frac{k_bT}{\mathcal{E}_{O \ VI}}}
\end{equation}
\begin{equation}
\label{Q(T)SIXII}
Q(T)_{Si \ XII}=Q_0^{Si \ XII} \frac{e^{-T_{cut}^{Si \ XII}/T}}{\sqrt{T}}\ln{\frac{k_bT}{\mathcal{E}_{Si \ XII}}}
\end{equation}

where $Q_0^{O \ VI}$ and $Q_0^{Si \ XII}$ are two normalization factors useful to match the emission
at the ionization equilibrium with the one prescribed by the CHIANTI database \citep{Landi2006},
$T_{cut}^{O \ VI}=139.000$ $K$ and $T_{cut}^{Si \ XII}=288.000$ $K$ are the temperature for O VI and Si XII respectively and
$\mathcal{E}_{O \ VI}=11.99$ $eV$ and $\mathcal{E}_{Si \ XII}=14.8$ $eV$ are the threshold energy respectively for O VI and Si XII.

\subsection{Radiative contribution \label{radiative}}
The radiative contribution is:

\begin{equation}
\label{IR}
I_r(n_i^Z,T,\vec{v})=I_0^r n_i^Z \sigma \mathcal{D}
\end{equation}

where $I_0^r$ is the intensity emitted from the Sun disk that reaches the height $r$
to be scattered by the coronal ions,
$\sigma$ is the cross section of the scattering,
and $\mathcal{D}$ is the Doppler dimming factor.
The radiative contribution is computed only for O VI, since it is negligible for Si XII.
$I_0^r$ is computed by:

\begin{equation}
\label{I0r}
I_0^r = I_0^{O \ VI \ 1032}2\pi\Biggl(1-\sqrt{1-\frac{R_{\odot}^2}{r^2}}\Biggr)
\end{equation}

where 
$I_0^{O \ VI \ 1032}=1.94$ $phot/(cm^2 \ s \ sr)$ is the disk intensity and the following function of $r$
is simply the dilution factor due to the distance from the disk.

$\sigma$ is computed (for Oxygen) by:
\begin{equation}
\label{sigmaOVI}
\sigma_{O \ VI} = \sigma_0^{O \ VI} \frac{v_{disk}^{O \ VI}}{\sqrt{\frac{3k_bT}{m_O}}}
\end{equation}

where $v_{disk}^{O \ VI}=30\times10^5$ $cm/s$ is the wideness of the Oxygen lines emitted from the disk,
and $m_O=2.67\times10^{-23}$ $g$ is the Oxygen mass.

$\mathcal{D}$ is computed (for Oxygen) by:
\begin{equation}
\label{DDOVI}
\mathcal{D}_{O \ VI} = (I_0^{O \ VI \ 1032}e^{-(\frac{v_{rad}}{v_{cut}^{O \ VI}})^2}
    +I_0^{Ly\beta} e^{-(\frac{(v_{rad}-v_0^{Ly\beta})}{v_{cut}^{H}})^2})/I_0^{O \ VI \ 1032}
\end{equation}

where the second term accounts for pumping of O VI by Ly$\beta$ \citep{Raymond2004} and
$I_0^{Ly\beta}=4.13\times10^{13}$ $phot/(cm^2 \ s \ sr)$ is the disk intensity for the Lyman $\beta$ line,
$v_0^{Ly\beta}=18\times10^{7}$ $cm/s$ is the distance between the center of the O VI 1032 $\AA$ line and the $Ly\beta$ line,
$v_{rad}$ is the radial velocity of the plasma and
$v_{cut}$ is the cutoff velocity for the Doppler dimming which is given by
the overlapping between the disk line profile and the coronal line profile.
For Oxygen and Hydrogen it is respectively:

\begin{equation}
\label{vcutOVI}
v_{cut}^{O \ VI}=\sqrt{v_{disk}^{O \ VI} + \frac{3k_bT}{m_O}}
\end{equation}
\begin{equation}
\label{vcutH}
v_{cut}^{H}=\sqrt{v_{disk}^{H} + \frac{3k_bT}{m_H}}
\end{equation}

where $v_{disk}^{H}=112\times10^5$ $cm/s$ is the width for Hydrogen lines emitted from the disk,
and $m_H=1.68\times10^{-24}$ $g$ is the Hydrogen mass.

%% file: appendix2.tex
\section{Analytical shock model \label{appendix2}}

From the Rankine-Hugoniot conditions for a planar adiabatic shock and our
radiation model presented in Appendix \ref{appendix1},
we develop an analytical model for the jump in the intensity of the emission lines due to the
shock.

First, we compute the jump in $n_e$, $n_i^Z$, $T$ and $\vec{v}$ as function of the Mach Number,
then we use the radiation model presented in Appendix \ref{appendix1}
to get the jump in the intensity of the emission.

Hereafter we indicate the pre-shock and post-shock quantities respectively with the subscript $0$ and $1$.
The jump in electron density, temperature and velocity due to the shock are \citep{Landau1966}:

\begin{equation}
\label{RHdens}
\frac{n_{e1}}{n_{e0}}=\frac{(\gamma+1)\mathcal{M}^2}{2+(\gamma-1)\mathcal{M}^2}
\end{equation}
\begin{equation}
\label{RHtemp}
\frac{T_1}{T_0}=\frac{2\gamma(\gamma-1)\mathcal{M}^4-(\gamma^2-6\gamma+1)\mathcal{M}^2-2(\gamma-1)}{(\gamma+1)^2\mathcal{M}^2}
\end{equation}
\begin{equation}
\label{RHvelocity}
v_1=2c_{s0}\frac{\mathcal{M}^2-1}{(\gamma+1)\mathcal{M}}
\end{equation}

where $\mathcal{M}$ is the Mach number and $c_s$ is the sound speed.

The ionization fraction is computed from the following approximated formula (Eq. (\ref{RHionsapp}))
in which pre-shock ionization equilibrium is assumed.

\begin{equation}
\label{RHionsapp}
f_i^Z=F_i^Z(T_1)+(F_i^Z(T_0)-F_i^Z(T_1))e^{-\frac{\tau}{\tau_{eq}}}
\end{equation}

where $f_i^Z$ is the ionization fraction for the ions $Z$ in the i-th ionization state,
$F_i^Z(T)$ is the equilibrium ionization fraction for the ions $Z$ in the i-th ionization state at temperature $T$,
$\tau$ is an estimation of the time elapsed from the shock front passage,
and $\tau_{eq}$ is the time scale needed to get ionization equilibrium which is given by:

\begin{equation}
\label{tauioniz}
\tau_{eq}=(n_e q_{eff})^{-1}
\end{equation}

Because a single rate dominates for both of the ions we are interested in, we can approximate $q_{eff}$
by the fastest rate involving the ion.

We can compute the ions density by:

\begin{equation}
\label{RHionsdensity}
n_i^Z=n_e A_Z^{Sun} f_i^Z
\end{equation}

where $A_Z^{Sun}$ is the solar abundance of the ion $Z$.

Through this computation we get the values for $n_e$,$n_i^Z$,$T$ and $\vec{v}$ for the post-shock region,
while we know all of them for the pre-shock region.
At this point we compute the jump in the emission lines from the radiation model presented in Appendix \ref{appendix1}.